# DNA looping in gene regulation: from the assembly of macromolecular complexes to the control of transcriptional noise

Jose M. G. Vilar[1] and Leonor Saiz[1,2]

[1]Computational Biology Center, Memorial Sloan-Kettering Cancer Center, 307 E. 63rd St., New York, NY 10021

[2]Center for Molecular Modeling and Department of Chemistry, University of Pennsylvania, 231 S. 34th St., Philadelphia, PA 19104

**Abstract**

The formation of DNA loops by proteins and protein complexes that bind at distal DNA sites plays a central role in many cellular processes, such as transcription, recombination, and replication. Here we review the basic thermodynamic concepts underlying the assembly of macromolecular complexes on looped DNA and the effects that this process has in the properties of gene regulation. Beyond the traditional view of DNA looping as a mechanism to increase the affinity of regulatory molecules for their cognate sites, recent developments indicate that DNA looping can also lead to the suppression of cell-to-cell variability, the control of transcriptional noise, and the activation of cooperative interactions on demand.

**Introduction**

Gene regulation relies to a great extent on proteins that bind to DNA, not only close to the genes they regulate but also at distal DNA sites that can be brought to the initiation of transcription region by looping the intervening DNA. This process, known as DNA looping, is widely used in gene regulation. It was first discovered in the *ara* operon of *E.*

*coli* [1], although it was already suspected to be present in eukaryotic enhancers [2], and since then it has been found in many other systems, such as the *gal*, *lac*, and *deo* operons in *E. Coli* [3-5], the lysogenic to lytic switch in phage λ [6], and the human β-goblin locus [7], to name just a few. In eukaryotes, multiple DNA binding sites that spread over long distances are involved in controlling the same localized DNA events. DNA looping is thus crucial to allow multiple proteins to affect the RNA polymerase in the promoter region. Enhancers, silencers, or mediators bound at distal DNA sites are then brought to form part of, affect, or interfere with the transcriptional complex. In prokaryotes, notwithstanding some exceptions [8,9], such a complexity in the signal integration machinery is not present and the role of DNA looping appears to be subtler, more like a fine-tuning of the underlying biochemistry.

Here we would like, on the one hand, to present the key thermodynamic concepts needed to understand the formation of the looped DNA-protein complexes and, on the other hand, to explore the effects that the formation of DNA loops has in gene regulation. We consider explicitly as prototype systems the *lac* operon [10] and phage λ [6], the two systems that led to the discovery of gene regulation. Both of them share many similarities, but they also exhibit crucial differences. In the *lac* operon, a single divalent protein, the *lac* repressor, can bind simultaneously to two DNA sites and induce the formation of a loop. In phage λ, the loop is not formed by a single protein but by a protein complex that is assembled on DNA when the loop forms. Phage λ is thus closer to eukaryotic transcription, where multiple proteins are assembled on the DNA to form the transcriptional complex with RNA polymerase.

**Increase in the local concentration and its paradoxes**

The *lac* operon is a clear example that illustrates the elegance of the functioning of biological systems and the subtlety upon which they are built. It consists of a regulatory domain and three genes required for the uptake and catabolism of lactose (Figure 1) [11]. The *lac* repressor can bind to the main operator $O_1$ and prevent the RNA polymerase from binding to the promoter and transcribing the genes. There are also two auxiliary operators, $O_2$ and $O_3$, where the repressor can also bind but not prevent transcription. Interestingly, elimination of either one auxiliary operator has only minor effects; yet simultaneous elimination of both of them reduces the repression level by about 100 times. The reason for this effect is that the *lac* repressor can bind simultaneously to two operators and loop the intervening DNA. Thus, the presence of $O_1$ and at least one auxiliary operator is enough to form DNA loops that substantially increase the repressor's ability to bind $O_1$. Binding to an auxiliary operator keeps the repressor close to $O_1$ and effectively increases its concentration around $O_1$. In this way, the efficiency of the binding to $O_1$ is greatly enhanced. However, both auxiliary operators are much weaker than $O_1$, as much as 10 ($O_2$) and 300 ($O_3$) times. For the increase in the local concentration to be a valid line of reasoning, the repressor needs to be bound to the auxiliary site in the first place. How is it then possible that a weaker operator can help the binding?

In phage λ there is also a related paradox. The lysogenic to lytic switch is controlled at two operators in the phage DNA, known as the left, $O_L$, and right, $O_R$, operators, which are 2.4kbp away from one another (Figure 1) [6,12]. Each of them has a tandem of three DNA motifs where the λ$c$I repressor dimers, can bind: $O_R1$, $O_R2$, and $O_R3$ for the right; and $O_L1$, $O_L2$, and $O_L3$ for the left operator. It was recently found that two λ$c$I dimers bound to $O_R1$ and $O_R2$ on the right operator can form an octamer with two λ$c$I dimers bound to $O_L1$ and $O_L2$ on the left operator by looping the intervening DNA [6,12-14]. The striking fact was not only that this fundamental result came more than forty years after the discovery of gene regulation in phage λ but also that the very idea of the increase of the local concentration just by itself can not be applied at all in this case. In the best scenario, current theories predict that the local concentration for such long loops would be increased by less than a factor 10 [15], which is well below the factor 1000 that is required to observe the formation of the octamers in solution [16]. The loop is thus too long to substantially increase the local concentration. How is it then possible that the loop forms when the octamer that ties it would not exist in such low concentrations?

The two counterintuitive examples we have just pointed out, namely, a weak site helping a strong site and a protein complex that would not exist in solution fastening a DNA loop, have a straightforward explanation when formulated in terms of the appropriate thermodynamic quantities.

**Free energies and the thermodynamic basis of regulated recruitment**

It is often assumed that cellular processes are like chemical reactions in an ideal well-stirred macroscopic reactor. The cell, however, is a small and crowded volume where many events take place at the same time. At the cellular level, the problem is not so much how to make two proteins interact but rather how to prevent them from interacting with all the other proteins they are not supposed to. Concentrations of the different molecular species are therefore kept low. To achieve specificity and affinity at the same time, cells have evolved mechanisms to bring molecules close to their interaction sites. It is now becoming more and more evident that this idea, referred to as regulated recruitment, is one of the unifying principles upon which living systems operate [17]. DNA looping is just one crucial example. In this section we will provide the basic thermodynamic ideas that will allow us to implement regulated recruitment in a quantitative way and apply it to DNA looping.

One key, perhaps the most important piece of information is that the free energy of binding, $\Delta G_{bind}$, can be decomposed into two main contributions:
$$\Delta G_{bind} = \Delta G_{pos} + \Delta G_{int}.$$
One, the interaction free energy $\Delta G_{int}$, arises from the interactions between the two molecules, such as electrostatic, hydrophobic, and Van der Waals interactions. The other, the positional free energy $\Delta G_{pos}$, results from positioning the molecules in the right place and orientation so that they can interact and it accounts, among other potential contributions, for the loss of translational and rotational entropy upon binding.

Let us consider in more detail the meaning of the positional free energy. If two molecules are to be bound, their positions have to be within a small volume of the order of the range of the interaction forces. The probability that one molecule is in this volume just by chance is given by the ratio of the volume of interaction, $V_{int}$, to the volume where the reaction takes place, $V_{reac}$. If there are $N$ molecules instead of one, the probability is scaled up accordingly, leading to the simple expression $P_{pos} = NV_{int}/V_{reac}$. Statistical thermodynamics [18,19] links this probability with its corresponding free energy, $\Delta G_{pos}$, through the relationship $P_{pos} \simeq e^{-\Delta G_{pos}/RT}$, where $R$ is the gas constant and $T$, the absolute temperature ($RT \simeq 0.6$ kcal/mol for typical experimental conditions). Equating both expressions for the positional probability and taking logarithms leads to $\Delta G_{pos} = -RT \ln(NV_{int}/V_{reac})$, which links the positional free energy with the concentration, $N/V_{reac}$, and a microscopic parameter of the binding, $V_{int}$.

For practical purposes, the expression for the positional free energy can be rewritten as
$$\Delta G_{pos} = \Delta G^o_{pos} - RT \ln[N]$$
by normalizing both volumes by the volume associated with one molecule at 1M concentration, $V_{mol} = 1.7$ nm$^3$, and by using the property that the logarithm of the product of two factors is the sum of the logarithm of each factor. The quantity $\Delta G^o_{pos} = -RT \ln(V_{int}/V_{mol})$ is the molar positional free energy and $[N] = NV_{mol}/V_{reac}$ is the concentration expressed in moles. In general, the free energy of binding depends on the concentrations of the different components through the positional free energy. Once it is known for a given concentration, $\Delta G^o_{pos}$ can easily be determined and the preceding expression gives the free energy values for any concentration. Free energies at 1M concentration are known as standard free energies and are usually labeled by the superscript $^o$.

Typical values of the positional free energy are $\Delta G^o_{pos} \approx 15$ kcal/mol [20]. This value indicates that if the free energy of interaction is zero, the probability that two molecules are as close as if they were bound is extremely small. Interaction forces are the ones that provide stability to the bound state. Even small values of binding free energies, such as $\Delta G_{bind} \approx -2$ kcal/mol, would imply considerably high interaction free energies, such as $\Delta G_{int} \approx -17$ kcal/mol. Now imagine that a univalent binding is transformed into a divalent one; i.e, that molecules originally having an interaction site now have two. The positional free energy would remain basically the same because it is associated with the molecule itself. The free energy of interaction, however, would be now twice as much, leading to $\Delta G_{bind} \approx -19$ kcal/mol in the preceding example. Similarly, a trivalent binding would lead to $\Delta G_{bind} \approx -36$ kcal/mol.

Such a high increase in affinity for multivalent binding, which is known in inorganic chemistry as the chelate effect [21], lies at the core of the assembly of the macromolecular complexes that implement regulated recruitment. Multiple binding domains also introduce additional complexity. In general, one should also consider the conformational free energy, $\Delta G_{conf}$, that accounts for the structural changes needed to accommodate multiple simultaneous interactions. The free energy of binding is thus the sum over all the contributions: $\Delta G_{bind} = \Delta G_{pos} + \sum \Delta G_{int} + \Delta G_{conf}$. In Figure 2a we show illustrative examples of the effects that different contributions have in the free energy of binding.

## Stability of the loop in the *lac* operon

These ideas can be applied to study the formation of the DNA loop between the main operator and the auxiliary operator in the *lac* operon [22]. The free energy of the looped state, $\Delta G_{O1 \sim O2} = \Delta G_{pos} + \Delta G_{int\,O1} + \Delta G_{int\,O2} + \Delta G_{loop}$, includes the positional free energy of one repressor, the interaction of each of the two repressor domains with $O_1$ and $O_2$, and the conformational free energy of forming the loop (Figure 2b). All these contributions must add up to the experimentally observed free energy (Table 1a), a substantially smaller quantity than the free energy of binding to just $O_1$, $\Delta G_{O1} = \Delta G_{pos} + \Delta G_{int\,O1}$ (Table 1a). Therefore, the extra interaction of the repressor with $O_2$ is able to compensate for the conformational free energy of forming a 401 bp DNA loop, which can be estimated to be $\Delta G_{loop} = \Delta G^o_{pos} + 8.4 \text{ kcal/mol}$ from the values of Table 1a and by combining different expressions for the free energy ($\Delta G_{loop} = \Delta G_{pos} + \Delta G_{O1 \sim O2} - (\Delta G_{O1} + \Delta G_{O2})$). Considering the apparent paradox of a weak site helping a strong one from this point of view, one realizes that the weak site is indeed weak with respect to the binding of a free repressor but strong compared to the conformational free energy of looping the DNA. In fact, the looped complex will be more stable than the repressor bound to $O_1$ when $\Delta G_{O2} < \Delta G_{pos} - \Delta G_{loop}$, which happens for standard free energies of binding to $O_2$ smaller than $-8.4 \text{ kcal/mol}$.

It is interesting to compare the energetics of the looped state with the free energy of the simultaneous binding of two repressors to $O_1$ and $O_2$, $\Delta G_{O1/O2} = \Delta G_{pos} + \Delta G_{int\,O1} + \Delta G_{int\,O2} + \Delta G_{pos}$. There is just one, yet crucial difference: the conformational free energy of closing the loop has been replaced by one positional free energy. At physiological *lac* repressor concentrations ($1.5 \times 10^{-8} \text{ M}$, about 10 repressors per cell), the positional free energy is $\Delta G_{pos} = \Delta G^o_{pos} + 11.3 \text{ kcal/mol}$ (recall $\Delta G_{pos} = \Delta G^o_{pos} - RT \ln[N]$), which indicates that the looped state is more stable. Only at concentrations higher than $8.3 \times 10^{-7} \text{ M}$, about 500 repressors per cell, would the simultaneous binding of two repressors dominate over the looped state.

The deconstruction procedure we have followed for the free energy has the advantage that the resulting contributions can be related to each other by combining them with the available experimental data for different experimental setups. From the *in vivo* measured activity of the *lac* operon [23] and its mathematical expression in terms of free energies [22], we have inferred the free energy of looping for different lengths of the DNA between operators (Table 1b). For intermediate lengths, within the range 150bp to 1.5kbp, the conformational free energy of looping nicely fits the theoretically predicted expression for an ideal flexible polymer $\Delta G_{loop}(l) = \Delta G_{loop}(l_0) + \alpha RT \ln(l/l_0)$, where $l$ is the length of the loop, $l_0$ a reference length, and $\alpha$ a constant. Intriguingly, theoretical estimates give α~ 2.25 [15,24], which is significantly different from the inferred *in vivo* value of α~1.25 (Table 1b).

**Assembly of macromolecular complexes in phage λ**

Phage λ represents a step forward in complexity. The loop is formed not by a single protein but by a protein complex that is assembled on the DNA as the loop forms. The free energy of the looped state with an assembled λcI octamer consists of different contributions, $\Delta G_{OR12 \sim OL12} = \Delta G_{intT} + \Delta G_{loop} + \Delta G_{OR12/OL12}$, which account for the interaction free energy between tetramers, the conformational free energy of forming the DNA loop, and the free energy of the tetramers bound to the right and left operators (Figure 2d). We can compare these contributions with those of the free energy of the octamer in solution, $\Delta G_{2 \to 8} = \Delta G_{intT} + \Delta G_{pos} + 2\Delta G_{2 \to 4}$, which account for interaction free energy between tetramers, the positional free energy, and free energy of the tetramers in solution, respectively (Figure 2c). The main difference between the formation of the octamer on DNA and in solution results from the formation of the constituent pair of tetramers, whose formation free energies at $0.5 \times 10^{-9} \text{M}$ are –4.2 and 10.6 kcal/mol, respectively (Tables 1c and 1d). The conformational free energy of forming the DNA loop, $\Delta G_{loop} = \Delta G_{pos}^o + 6.8 \text{ kcal/mol}$, can be obtained from the preceding equations for $\Delta G_{OR12 \sim OL12}$ and $\Delta G_{2 \to 8}$ and Tables 1c and 1d. It is smaller than the positional free energy required to bringing the tetramers together in solution, $\Delta G_{pos} = \Delta G_{pos}^o + 12.8 \text{ kcal/mol}$. Therefore, DNA acts as a scaffold for the formation of tetramers and helps also the tetramers to encounter each other, which explains why an octamer that does not exist in solution is able to fasten a DNA loop in phage λ.

Interestingly, once the tetramers are formed, the interaction free energy between them is barely able to compensate for the conformational free energy cost of closing the loop: $\Delta G_{intT} + \Delta G_{loop} = -0.5 \text{ kcal/mol}$. The octamer is formed on looped DNA, but it is present at most 70% of the time. In order to further stabilize the loop, binding of λcI dimers to $O_R 3$ and $O_L 3$ is required. In this case, the free energy of forming the loop from the non-looped complex is $\Delta G_{intD} + \Delta G_{intT} + \Delta G_{loop} = -3.5$ kcal/mol, where $\Delta G_{intD}$ is the interaction free energy between dimers bound to $O_R 3$ and $O_L 3$ (Figure 2d). Such contribution to the free energy allows the loop to be present up to 99.7% of the time.

## Effects of DNA looping in gene regulation

The thermodynamic approach we have presented provides a straightforward method to obtain the free energy of the assembly of macromolecular complexes from the different contributions of their components. Remarkably, there are potentially much more complexes than components. Therefore, measuring the free energy of a selected set of complexes can be sufficient to thermodynamically characterize all the components and their mutual interactions. The resulting information can in turn be used to obtain the free energies of all the possible complexes. Knowing the free energies of all the complexes, or equivalently, all the possible configurations in which the components can be arranged, allows statistical thermodynamics to make quantitative predictions about the probability of finding different configurations [18] and the resulting effects in gene regulation [19].

The widespread view in the field of gene regulation is that DNA looping is just a mechanism to increase the binding of regulatory molecules to their corresponding DNA binding sites [3]. In fact, the thermodynamic approach we have discussed shows how such increase is achieved in representative instances. DNA looping, however, is intrinsically different from other common mechanisms that could be used to increase the affinity of a molecule for its cognate site. We discuss below three crucial effects that DNA looping has on gene regulation, namely, the suppression of cell-to-cell variability, the control of transcriptional noise, and the supply of cooperativity on demand.

## Cell-to-cell variability

The numbers of different molecular species are expected to differ from cell to cell, especially when they are as low as tens or hundreds. Such differences can even make genetically identical cells behave differently under the very same conditions [25,26].

Therefore, it is important to understand how the molecular and behavioral variability are connected through the different cellular processes.

Let us focus again on the *lac* operon. A useful quantity to measure the strength of a repressor is the repression level, which is defined as the maximum transcription divided by the actual transcription. Because transcription takes place when the repressor is not bound to the main operator $O_1$, the actual transcription rate is the maximum rate times the probability for the main operator to be free. Thus 0%, 95%, and 99.95% occupancy of $O_1$ by the *lac* repressor results in a repression level of 1, 21, and 2001; and in the production of 6000, 300, and 3 β-galactosidase molecules (the product of the *lacZ* gene) per cell each hour, respectively [10].

The parameters that characterize different mechanisms of transcription regulation can be chosen so that the repression level is the same for all of them at a given number of repressors per cell. However, differences are expected to arise as soon as the numbers of repressors change. We consider explicitly the repression level for three alternative mechanisms: two operators and DNA looping, as in the *lac operon* without $O_3$; a tandem of two identical $O_1$ operators without DNA looping, so that both operators have to be free for transcription to occur; and a single, but stronger operator so that the repression level is the same as for the looping case when the number of the repressors per cell is 10. Strikingly, the repression level for the looping case is a convex function of the number of repressors (Figure 3a), which leads to a low sensitivity to changes in the number of repressors. The other two cases exhibit linear and quadratic relationships between the change in the repression level and the number of repressors. The implications of these dependences extend up to the cell population level. The low sensitivity obtained for DNA looping can be used to achieve fairly constant repression levels among cells in a population (Figure 3c) irrespective of the fluctuations of the numbers of regulatory molecules (Figure 3b). In contrast, using a single operator just propagates the fluctuations proportionally and two operators without DNA looping can lead to an amplification of the underlying molecular variability (Figures 3b and 3c).

**Transcriptional noise**

Differences in the numbers of regulatory molecules are not the only source of cellular variability. The intrinsic stochasticity of cellular processes, usually referred to as noise, also plays an important role [26]. When the average production of proteins is as low as 10 per hour, stochastic effects can become relevant for the behavior of the system [27-32]. Figures 3d and 3e show the typical time courses and the histograms of the number of molecules produced from operons regulated with DNA looping and with just a single binding site. The parameters of the system are such that the single operator cases have the same repression level and average transcription rate as for DNA looping. Nevertheless, as the figure clearly illustrates, these mechanisms are not equivalent, not even when the numbers of regulatory molecules are kept constant. In particular, if a lower repressor-operator dissociation rate constant is used to increase the repression level up to that obtained with DNA looping, fluctuations are greatly enhanced. In contrast, using a higher association rate constant for that purpose will keep the fluctuations as low as for the looping case.

The reason for these differences is a matter of time scales. If transcription switches slowly between active and inactive, there are long periods of time in which proteins are produced constantly and long periods without any production. Therefore, the number of molecules fluctuates strongly between high and low values. In contrast, if the switching is very fast, the production is in the form of short and frequent bursts. This lack of long periods of time with either full or null production gives a narrower distribution of the number of molecules. DNA looping naturally introduces a fast time scale: the time for the repressor to be recaptured by the main operator before unbinding the auxiliary operator is much shorter than the time needed by a repressor in solution [22]. DNA properties are therefore important in controlling transcriptional noise. Conspicuously, it has recently been observed that DNA can cyclize at unusually high rates [33-35].

Looping and a higher association rate constant might seem to provide equivalent mechanisms regarding fluctuations. There are, however, certain limits for the values that the rate constants can achieve. The theoretical limit for the association rate constant of diffusion-limited reactions is $k_a \simeq 10^9 M^{-1} s^{-1}$ [36]. To reduce the fluctuations by increasing the association rate constant, the diffusion limit would have to be surpassed, which does not seem to be the case for the *lac* repressor [37,38]. DNA looping consequently provides the cell with a mechanism to circumvent the physical constraints imposed by diffusion-limited reactions.

## Cooperativity on demand

The logics of λcI regulation is to activate its own transcription at the $P_{RM}$ promoter and repress the cro gene at the $P_R$ promoter (Figure 1) [6]. λcI dimers bind cooperatively to $O_R 1$ and $O_R 2$ to form a tetramer on the DNA. Once the concentration of λcI is sufficiently high, $O_R 3$ gets occupied and its transcription at the $P_{RM}$ promoter is turned off. For a long time, one of the main puzzles in the regulation of phage λ was that the strength of $O_R 3$ was too weak for it to be occupied at physiological λcI concentrations [39].

It is now clear that $O_R 3$ is occupied at physiological concentrations because of the effects of DNA looping [6,12,13]. In Figures 3f and 3g we show the activity of the $P_R$ and $P_{RM}$ promoters as a function of λcI dimer concentration, [λcI$_2$], for different values of the free energy of looping [14]. As the free energy of looping decreases, $O_R 3$ becomes more occupied by λcI dimers (Figure 3f). Thus DNA looping provides the cooperativity needed for the occupation of $O_R 3$.

Unexpectedly, if the free energy of looping is decreased further beyond wild type levels, the qualitative behavior resembles that of the case when there is no DNA looping at all. Let us consider this key point in more detail. For the phage λ switch to function properly, repression of the *cro* gene and activation of the *λcI* gene should occur simultaneously. This is accomplished by the formation of λcI tetramers at the $O_R$ operator. In addition, as soon as λcI concentration is sufficiently high, its production should be turned off to allow effective RecA-mediated switch to the lytic state [6]. If the free energy of looping is too high, DNA cannot loop and *λcI* is repressed only for concentrations well above those of *λcI* activation and *cro* repression. If the free energy of looping is too small, DNA readily loops and *λcI* activation and *cro* repression occur at a much lower concentration than that required for *λcI* repression. Therefore, both high and low conformational free energies of looping lead to the same qualitative behavior. The main difference is just a shift in concentrations at which different regulatory events happen (Figures 3f and 3g). Only an intermediate range of free energies of looping around the *in vivo* value will give the adequate behavior. So far, all these properties of DNA looping have not been considered in quantitative approaches [40] to analyze the remarkable stability of the switch [41].

DNA looping is thus crucial for the high sensitivity of the activation and repression of the $P_{RM}$ promoter. The fact that the DNA loop can open and close upon small changes of λcI dimer concentration allows DNA looping to sharply switch on the cooperativity that it provides to the binding of the λcI dimers to $O_R3$ and $O_L3$, and thus maintain a tightly regulated concentration of λcI dimers in the lysogenic state. Cooperativity between $O_R3$ and $O_L3$ binding is therefore only present when the loop is formed by the λcI octamer. A similar pattern of induced cooperativity is also observed in RXR, a nuclear hormone receptor [42]. In its tetrameric form, RXR has two DNA binding domains and can loop DNA to bring transcription factors close to the promoter region. Retinoic acid controls whether or not the loop is formed by preventing the assembly of the tetrameric complex from the constituent dimers, which also bind DNA.

## Conclusions

Regulation systems have evolved constrained by the intrinsic molecular nature of the cell. Cells are densely packed with thousands of different molecular species and their function is built upon molecular events that are inherently stochastic. Integration of such multiplicity of components into a functional unit must be able to balance a series of factors all of which might not be attainable at the same time. If concentrations of the different molecular species are kept low to prevent non-specific interactions, not only the binding to the specific sites is decreased but also fluctuations are expected to become important. DNA looping is a mechanism that can be used to increase specificity and affinity simultaneously, and, at the same time, to control the intrinsic stochasticity of cellular processes. In particular, it can buffer molecular variability to produce phenotypically homogeneous populations, decrease the transcriptional noise, and allow cooperative interactions to take place on demand, as required from the cellular context.

It is clear that the effects of DNA looping in gene regulation cannot be fully understood just in terms of increased local concentrations of the regulatory molecules. DNA looping relies on key thermodynamic quantities that extend beyond the macroscopic theory of chemical reactions. Explicitly, the free energy of each binding molecule can be decomposed into a single unfavorable positional free energy and multiple, potentially favorable interaction free energies (one per interacting binding domain). Such deconstruction provides a starting point to characterize and predict the collective properties of macromolecular complexes, such as looped DNA-protein complexes, in terms of the properties of its constituent elements.

Uncovering how different molecular mechanism determine the cellular behavior is of fundamental importance for understanding both naturally occurring [17,43] and artificially designed cellular systems [44].

## Acknowledgments

We are indebted to Nicolas Buchler, Calin Guet, Stanislas Leibler, Mark Ptashne, Miguel Rubi, Nikolaus Schultz, Wenying Shou, and Jon Widom for comments and discussions.

# Figure and table captions

## Figure 1

Scheme of the different operator positions on the DNA for the *lac* operon and phage λ. *(top)* Location of the main (O1) and auxiliary operators (O2 and O3) of the *lac* operon. Binding of the lac repressor to O1 represses transcription of the *lacZ*, *lacY*, and *lacA* genes. *(bottom)* Location of the right ($O_R$) and left operators ($O_L$). Binding of the λcI dimer to $O_R2$ activates transcription of its own gene. Binding of λcI dimers to $O_R1$ and to $O_R3$ prevents transcription of *cro* and *λcI* genes, respectively.

## Figure 2

Examples of different contributions to the free energy of binding. *(a)* Illustrative imaginary situations for zero, one, and two interaction domains. The different contributions to the free energy of binding, $G_{bind}$, are a positional free energy of 15 kcal/mol, an interaction free energy of -17 kcal/mol for each interaction site, and a conformational free energy of 10 kcal/mol. *(b)* Energetics of the formation of the DNA loop-lac repressor complex. Contributions to the free energy of the complex formed by the divalent lac repressor (shown in gray) when its two DNA binding domains interact simultaneously with the two operators O1 and O2 by looping the intervening DNA: $G_{O1\sim O2} = G_{pos} + G_{intO1} + G_{intO2} + G_{loop}$. The positional free energy term $G_{pos}$ accounts for the free energy necessary to bring the *lac* repressor (shown in white with dashed contour lines) to the appropriate position and orientation (shown in gray with continuous contour lines) so it can bind the operators. Two interaction free energy terms account for the interaction between the two domains of the lac repressor with the O1 ($G_{intO1}$) and O2 ($G_{intO2}$) operators, respectively. The last contribution to the free energy of the DNA loop-lac repressor complex accounts for the conformational free energy cost of looping the DNA between the two operators ($G_{loop}$). *(c)* Energetics of the formation of λcI octamers from tetramers in solution. At physiological conditions, λcI exists in solution as a dimer (shown in white with dashed contour lines) and as a monomer (not shown). Dimers can oligomerize to form tetramers and two tetramers can form an octamer. The free energy of forming λcI octamers (shown in gray as the assembly of four dimers) from the tetramers in solution (shown in white with dashed contours as the assembly of two dimers) is given by $G_{4 \to 8} = G_{pos} + G_{intT}$. $G_{pos}$ is the positional free energy of bringing the tetramer from solution (shown in white with dashed contour lines) to form part of the octameric complex (shown in gray with continuous contour lines) and $G_{intT}$ is the interaction free energy between the two tetramers that form the octamer. *(d)* Energetics of the λcI-DNA loop formation. $G_{intT}$ is the interaction free energy between λcI tetramers bound to OR1,2 and to OL1,2; $G_{intD}$ is the interaction free energy between dimers bound to OR3 and OL3; and $G_{loop}$ is the conformational free energy of looping the intervening DNA. The free energy of forming the λcI-DNA loop complex from a non-looped conformation with λcI dimers (shown in gray) bound to all the operator sites is given by $G_{R123 \sim L123} - G_{R123/L123} = G_{intT} + G_{intD} + G_{loop}$.

## Table 1

Representative free energies (in kcal/mol) of binding and looping states for the *lac* operon and phage λ at 1M (standard conditions) and at the physiological concentrations of $1.5 \times 10^{-8}$ M (*lac* operon) and $0.5 \times 10^{-9}$ M (phage λ). Whenever DNA looping can be present, the tilde (~) and slash (/) symbols in the labels of the states indicate looped and non-looped DNA, respectively. *(a)* Binding of the *lac* repressor to different operators. The free energy of binding was inferred in reference [22] from the experimental data of reference [23]. *(b)* Free energy of looping as a function of the distance between operators for the *lac* operon. The symbols correspond to the estimation obtained from the experimental data [23] for the repression level, *R*, and the mathematical expression that links it to the free energy: $R = 1 + e^{-\Delta G_{O1}/RT}([N] + e^{-(\Delta G_{loop}(l) - \Delta G_{pos}^o)/RT})$ [22]. The continuous line corresponds to the displayed equation. (c) Binding of λcI dimers to different right operators (top section), left operators (middle section), and left and right operators at the same time with looped and non-looped DNA conformations (bottom section) inferred in reference [14] from diverse experimental sources. (d) Free energies of the formation of λcI octamers ($2 \rightarrow 8$) and tetramers ($2 \rightarrow 4$) from dimers in solution; and free energy of the formation of λcI octamers ($4 \rightarrow 8$) from the tetramers in solution [16].

**Figure 3**

Effects of DNA looping in gene regulation. <u>Cell-to-cell variability</u>: *(a)* Computed repression level as a function of the number of repressors *N* for two operators and DNA looping, as in the *lac* operon without O3 (black line) [22]; a single operator, as in the *lac* operon without O3 and O2 (red line) [22]; and two identical operators that overlap with the promoter (blue line). The different mechanisms are schematically represented next to each line. The curves corresponding to the continuous red and blue lines were computed for parameters chosen to obtain the same repression level as for the looping case for a number of repressors *N* =10. This value of *N* corresponds to the wild type average number of *lac* repressors per cell. Assuming a Poisson distribution for fluctuations in the number of repressors, such as that shown in *(b)*, the corresponding distribution of repression levels obtained for the three cases considered (looping in black, single stronger operator in red, and two proximal operators controlling RNA polymerase binding in blue) are plotted in *(c)*. <u>Transcriptional noise</u>: *(d)* Time courses and *(e)* histograms of the number of molecules produced from operons regulated with *(top)* and without looping. When the repressor is not bound to $O_1$, β-galactosidase molecules, encoded by the *lacZ* gene, are randomly produced at a rate of 13 per second and degraded with a characteristic life-time of 30 min. *(top)* Regulation with looping (rates are the same as in Ref. [22]). *(center)* Regulation with a single (but stronger) operator with a dissociation rate constant 0.021 times smaller, chosen so that the repression level is the same as for the looping case. *(bottom)* Same situation as in *(center)* but with the same dissociation rate constant of *(top)* and an association rate constant 47 times larger, which is beyond the limit of diffusion-limited reactions. <u>Cooperativity on demand</u>: Production of LacZ from the *(f)* $P_{RM}$ promoter that transcribes the λ*cI* gene and from the *(g)* $P_R$ promoter that transcribes the *cro* gene as a function of the concentration of λcI dimers, $[\lambda cI_2]$, computed similarly as in reference [14]. Regulation by looping (black continuous lines) is compared to regulation when the looped DNA conformation cannot be formed because of a too high free energy of looping (red continuous lines) and when the DNA is trapped in a looped conformation because of a too low free energy of looping (blue continuous line). The free energies of looping for each case are $G_{loop} = -G_{intT} - 0.5 \text{kcal/mol}$, $G_{loop} = -G_{intT} + 5 \text{kcal/mol}$, and $G_{loop} = -G_{intT} - 5 \text{kcal/mol}$, respectively. The latter case can be adequately scaled for comparison with the other two cases, as shown by the blue dashed lines, by plotting the production of LacZ as a function of $[\lambda cI_2] e^{0.5 \Delta G_{intD}/RT}$ instead of $[\lambda cI_2]$. Each curve is labeled with its corresponding value of $G_{loop} + G_{intT}$. The symbol -5(s) labels the scaled curves.

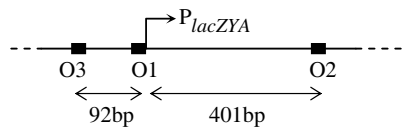

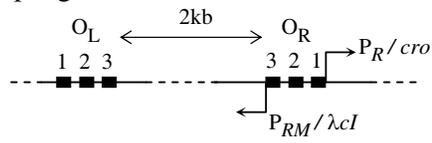

Figure 1

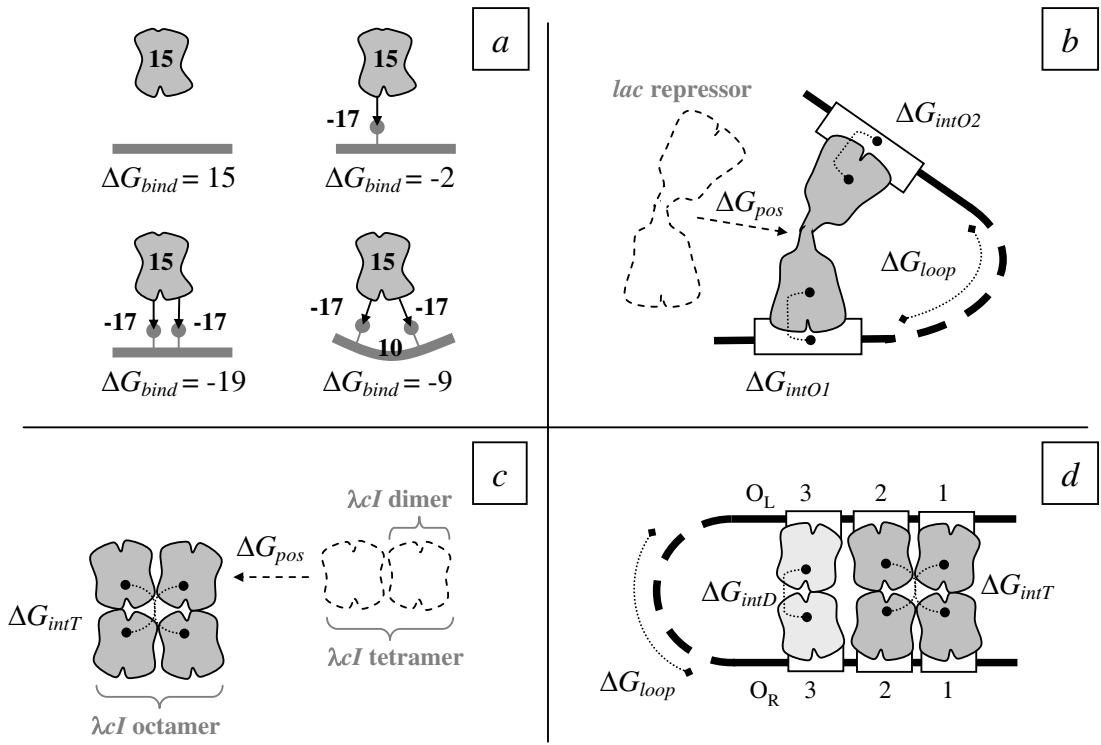

Figure 2

| lacR / state $i$ | $\Delta G_i^o$ | $\Delta G_i^{1.5\times 10^{-8}\,M}$ | a |
|---|---|---|---|
| $O1$ | −13.7 | −2.9 | |
| $O2$ | −11.8 | −1.0 | |
| $O1/O2$ | −25.5 | −4.0 | |
| $O1 \sim O2$ | −17.1 | −6.3 | |

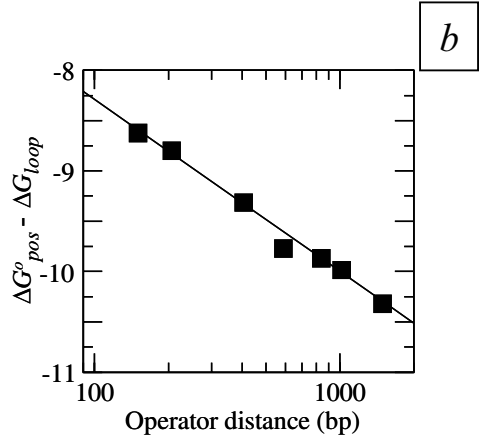

b

$$\Delta G_{loop} = \Delta G_{pos}^o + 4.9 + 1.24 RT \ln(l)$$

| λcI / state $i$ | $\Delta G_i^o$ | $\Delta G_i^{0.5\times 10^{-9}\,M}$ | c |
|---|---|---|---|
| $O_R 1$ | −13.2 | −0.4 | |
| $O_R 2$ | −10.7 | 2.1 | |
| $O_R 3$ | −10.2 | 2.6 | |
| $O_R 1 O_R 2$ | −26.9 | −1.3 | |
| $O_R 1 O_R 2 O_R 3$ | −37.1 | 1.2 | |
| $O_L 1$ | −13.8 | −1.0 | |
| $O_L 2$ | −12.1 | 0.7 | |
| $O_L 3$ | −12.4 | 0.4 | |
| $O_L 1 O_L 2$ | −28.4 | −2.8 | |
| $O_L 1 O_L 2 O_L 3$ | −40.8 | −2.5 | |
| $R12 \sim L12$ | −55.8 | −4.7 | |
| $R12 / L12$ | −55.3 | −4.2 | |
| $R123 \sim L123$ | −81.4 | −4.7 | |
| $R123 / L123$ | −77.9 | −1.2 | |

| λcI / process $i$ | $\Delta G_i^o$ | $\Delta G_i^{0.5\times 10^{-9}\,M}$ | d |
|---|---|---|---|
| $2 \to 4$ | −7.5 | 5.3 | |
| $2 \to 8$ | −22.3 | 16.0 | |
| $4 \to 8$ | −8.0 | 4.8 | |

Table 1

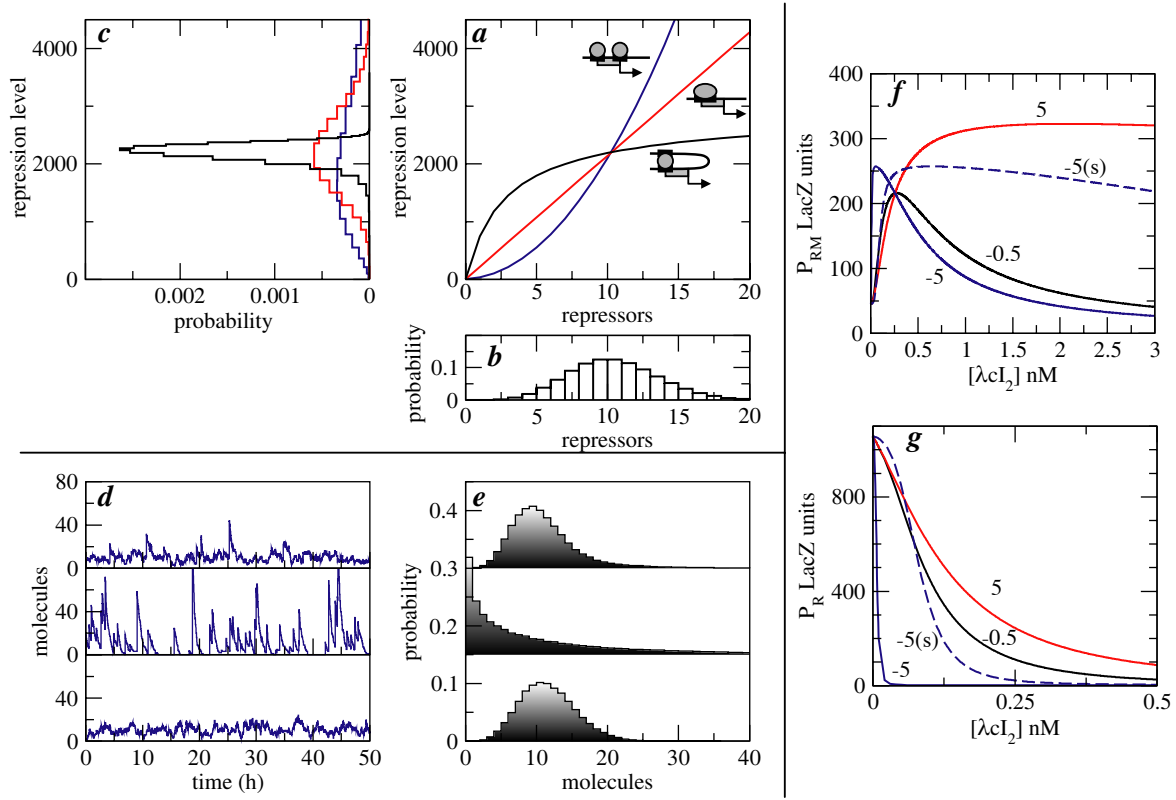

Figure 3